\documentclass[amsmath,amssymb,prd,twocolumn,final,superscriptaddress]{revtex4-2}
\usepackage{dcolumn}
\pdfoutput=1
\usepackage[english]{babel}
\usepackage[utf8]{inputenc}
\usepackage[T1]{fontenc}
\usepackage{amsthm}
\usepackage{amsfonts}
\usepackage{url}
\usepackage{tensor}
\usepackage[colorlinks = true,
            linkcolor = blue, 
            linkcolor = blue,
            urlcolor  = blue,
            citecolor = blue,
            anchorcolor = blue]{hyperref}
\usepackage{graphicx}
\usepackage{mathrsfs}
\usepackage{enumerate}
\usepackage{caption}
\usepackage{autonum}
\makeatletter
\newcommand{\forceeqnum}[1]{\autonum@markLabelAsReferenced{#1}}
\makeatother
\usepackage{orcidlink}

\newcounter{mnotecount}[section]

\newcommand{\Ldot}{\!\raisebox{0.5ex}{\scalebox{1}{$\cdot$}}}

\newcommand{\beq}{\begin{eqnarray}}
\newcommand{\eeq}{\end{eqnarray}}
\newcommand{\ben}{\begin{eqnarray*}}
\newcommand{\een}{\end{eqnarray*}}

\newtheorem*{theorem*}{Theorem}
\theoremstyle{definition}

\DeclareMathOperator{\sech}{sech}

\newcommand\ringring[1]{%
  {
   \mathop{\kern0pt #1}\limits^{
     \vbox to-1.85ex{
       \kern-2ex 
       \hbox to 0pt{\hss\normalfont\kern.1em \r{}\kern-.45em \r{}\hss}
       \vss 
     }   }
  }
}

\makeatletter
\def\frontmatter@title@above{\addvspace{6\p@}}
\def\frontmatter@preabstractspace{11.5\p@}
\def\frontmatter@postabstractspace{12.5\p@}
\makeatother

\begin{document}
\title{Backreaction of Halilsoy and Chandrasekhar waves}
\author{Sebastian J. Szybka\,\orcidlink{0000-0003-3648-9285}}
\affiliation{Astronomical Observatory, Jagiellonian University}
\author{Adam A. Zychowicz\,\orcidlink{0009-0004-2544-0632}}
\affiliation{Astronomical Observatory, Jagiellonian University}
\affiliation{Doctoral School of Exact and Natural Sciences, Jagiellonian University}
\author{Dominika Hunik\,\orcidlink{0000-0002-1804-6897}}
\affiliation{Astronomical Observatory, Jagiellonian University}

%\date{}
\begin{abstract}
We calculate the high-frequency limit of the Halilsoy and Chandrasekhar standing gravitational wave solutions. We show that the backreaction effect is the same for these classes of solutions and we analyze the causal structure of the effective spacetime. In addition, we rederive both classes of solutions without referring to the Ernst equation and generation techniques.
\end{abstract}
\maketitle{}

\section{Introduction}

The first detection of gravitational waves opened a new era in astronomy by freeing observers from the limitations of electromagnetic radiation. On the theory side, the linearized Einstein equations are well understood, but the nonlinear regime is still a challenge. The Einstein equations for the plane waves can be reduced to the Laplace equation, which is linear; consequently,  plane waves traveling along the same direction do not interact. The simplest nontrivial toy models are cylindrically symmetric. This kind of models proved to be a useful tool in general relativity \cite{EinsteinRosen37,PiranSafier:1985,marder,bpulse}. 

Our study is focused on two complementary questions. The first one, raised by Hans Stephani \cite{Stephani:2003}, concerns the definition and properties of standing gravitational waves. In cylindrical symmetry, two different classes of standing-wave spacetimes are known. The Halilsoy spacetimes \cite{Halilsoy:1988} are the exact solutions to the vacuum Einstein equations. By contrast, the Chandrasekhar solutions \cite{chandra} are obtained under more restrictive assumptions and the essential part of the Einstein equations has to be solved numerically. The comparative study of these two classes sheds light on the subtleties regarding definition and properties of standing gravitational waves. The aim of the recent paper \cite{NikielSzybka:2025}, coauthored by one of us, was to clarify differences and similarities between these standing gravitational waves in the weak-field limit. It has been shown there that these solutions in this approximation and far away from the symmetry center correspond to polarization standing gravitational waves. The differences between these two solutions are related to the behavior of their polarizations. In this paper, which is an extension of the article mentioned above \cite{NikielSzybka:2025}, we rederive both solutions without resorting to the Ernst equation and solution-generating techniques. Our derivation reveals the similarities and differences between these two classes of solutions. We also compare both classes of solutions in the high-frequency limit. To this end, we apply the Green--Wald framework \cite{greenwald}, which allows us to compare backreaction effects. 

The problem of backreaction---how short-wavelength structure influences the large-scale geometry---is central to general relativity. Therefore, the high-frequency limits presented in this paper are interesting in their own right, not only in the context of standing gravitational waves. The particular version of the backreaction problem investigated by us has been studied by many authors before and concerns the backreaction of gravitational radiation  \cite{isaacson1,isaacson2,Maccallum:1973gf,burnett,Podolsky:2003bm,greenwald}. The gravitational waves carry energy. If the scales considered are much larger than the wavelength of gravitational waves, then the energy of the waves should be treated as a source of the gravitational field. The backreaction problem, in this context, is to determine the effective spacetime that accounts for the energy of the waves. The related question is of great importance in cosmology, where the energy of the stochastic gravitational background is dominated by the energy associated with small-scale inhomogeneities in the distribution of matter. 

Over the years, many attempts were made to solve the backreaction problem. In the context of vacuum spacetimes, Burnett defined the so-called weak limit, which makes it possible to determine the effective energy-momentum tensor in the high-frequency limit \cite{burnett}. This tensor describes the energy content of the small spacetime ripples and allows one to determine the corresponding effective spacetime. Green and Wald \cite{greenwald} extended Burnett's approach to nonvacuum spacetimes. Burnett conjectured that all spacetimes obtained as a high-frequency limit must solve the Einstein-massless Vlasov system. His hypothesis was partially proven under additional assumptions \cite{Huneau:2025mry}. It is also consistent with all examples provided so far where the majority of known solutions are special cases of the Einstein-massless Vlasov spacetimes, namely, the spacetimes sourced by null dust. The converse conjecture, that all solutions to the Einstein-massless Vlasov system can be obtained as a high-frequency limit is also being actively investigated \cite{Huneau:2024}. The Burnett and Green--Wald formalisms are based on the assumption of the existence of a one-parameter family of solutions to the Einstein equations. In order to calculate the high-frequency limit, we construct such an analytic one-parameter family of solutions within the Halilsoy class. Similarly, we construct an analytic one-parameter family of spacetimes that approximates the one-parameter family of solutions within the Chandrasekhar class. We show that both families of solutions can be approximated in the high-frequency limit by the same effective spacetime. It corresponds to the Morgan solution \cite{morgan}, which solves the Einstein equations with a superposition of null dusts as sources. We study its properties. Our results provide noncosmological examples of solutions that satisfy the assumptions of the Burnett and Green--Wald frameworks and enrich the still short list of known high-frequency limits. This list starts with the pioneering work of Choquet-Bruhat \cite{choquet}---for a review of known examples see the article \cite{Huneau:2024lrb}. To sum up, the high-frequency limits presented in this paper are of interest in the context of studies of the backreaction \cite{greenwald2,ourwm,ers,Huneau:2017led,Huneau:2024lrb,Huneau:2024,Luk:2020pyn,Huneau:2025mry}. 

Since solutions we consider are cylindrically symmetric, they are particularly useful as mathematical toy models for standing gravitational waves and backreaction. However, cylindrically symmetric standing gravitational waves also have been discussed in experimental setups \cite{RevModPhys.52.341,GrishchukSazhin1975}.

\section{Cylindrical standing gravitational waves}

\subsection{Einstein equations}

Let us consider the cylindrically symmetric metric in the coordinates $(t,\rho,\phi,z)$ adapted to the Killing fields $\partial_\phi$ and $\partial_z$:
\begin{equation}\label{metricH}
g=e^{2(\gamma-\psi)}\left(-dt^2+d\rho^2\right)+\rho^2e^{-2\psi}d\phi^2+e^{2\psi}(dz+\omega d\phi)^2\;,
\end{equation}
where $\rho>0$, $-\infty<t,z<\infty$, $0\leq\phi<2\pi$ and \mbox{$\psi=\phi(t,\rho)$}, $\gamma=\gamma(t,\rho)$, $\omega=\omega(t,\rho)$.

The Einstein equations $G_{\alpha\beta}=0$ reduce in vacuum to the vanishing of the Ricci tensor $R_{\alpha\beta}=0$. In axial symmetry they can be presented in a concise form as the Ernst equation. Using generation techniques one may derive Halilsoy \cite{Halilsoy:1988} and Chandrasekhar \cite{chandra} solutions. We follow a more direct approach.  From $R^0_{\;1}=0$ and $G^0_{\;0}=0$ (which coincides with $R^0_{\;0}-R^1_{\;1}=0$) we obtain the equations
\begin{equation}\label{Einstein1}
\begin{split}
\dot{\gamma}=&2\rho \dot\psi\psi'+\frac{e^{4\psi}}{2\rho}\dot\omega\omega'\;,\\
	\gamma'=&\rho\left(\dot\psi^2+{\psi'}^2\right)+\frac{e^{4\psi}}{4\rho}\left(\dot\omega^2+{\omega'}^2\right)\;,
\end{split}
\end{equation}
where the dot and prime denote the derivatives in $t$ and $\rho$, respectively. If $\psi$ and $\omega$ are known, the metric function $\gamma$, which corresponds to C-energy \cite{thorne}, can be determined via quadratures. The equations $R^2_{\;2}-\omega R^2_{\;3}=0$ and \mbox{$R^2_{\;3}=0$} give
%
%\begin{equation}
\begin{align}
	-\ddot\psi+\psi'/\rho+\psi''=&\frac{e^{4\psi}}{2\rho^2}\left(\omega'^2-\dot\omega^2\right)\;,\label{Einstein2a}\\
	-\ddot\omega-\omega'/\rho+\omega''=&4\left(\dot\psi\dot\omega-\psi'\omega'\right)\;.\label{Einstein2b}\\
\end{align}
%\end{equation}
%
For vacuum, we have $R=0$, thus $G^0_{\;0}=0\implies R^0_{\;0}=0$. Therefore Eqs.\ \eqref{Einstein1}--\eqref{Einstein2b} imply $$R^0_{\;0}=R^1_{\;1}=R^0_{\;1}=R^2_{\;2}=R^2_{\;3}=0\;.$$ Moreover, $R=0$ together with the equation above implies \mbox{$R^3_{\;3}=0$.} The remaining components of the Ricci tensor vanish identically or as a consequence of the symmetry of this tensor.

\subsection{Solutions}

The left-hand side of Eq.\ \eqref{Einstein2a} corresponds to the d'Alembert operator acting on $\psi$ (in a flat spacetime, in cylindrical coordinates). If the right-hand side in \eqref{Einstein2b} vanishes, then this equation becomes a special case of the Euler--Poisson--Darboux equation. If $\omega=0$, then the solutions to the Einstein equations \eqref{Einstein1}--\eqref{Einstein2b} belong to the Einstein--Rosen family \cite{EinsteinRosen37}. Equations \eqref{Einstein2a}, \eqref{Einstein2b} constitute the essential part of the system and they can be written as the Ernst equation \cite{Ernst:1967wx,Ernst:1967by}. 

In order to better understand differences between the Halilsoy and the Chandrasekhar solutions, we derive both solutions in a more direct way, without referring to the Ernst equation and generation techniques.

We adopt a natural simplifying assumption---the metric function $\omega$, which controls polarization of the solutions is strictly periodic in time and satisfies the harmonic oscillator equation 
\begin{equation}\label{harmonicO}
	\ddot\omega+\frac{1}{\lambda^2}\omega=\frac{Y}{\lambda}\;,
\end{equation}
where $\lambda>0$ is a constant and $Y=Y(\rho)$ is an arbitrary dimensionless function of $\rho$ ($\omega$ has dimension of length). 

Equation \eqref{Einstein2b} can be written as
\begin{equation}\label{Einstein3b}
	\frac{1}{\lambda^2}\omega-\omega'/\rho+\omega''=\frac{Y}{\lambda}+4\left(\dot\psi\dot\omega-\psi'\omega'\right)\;.
\end{equation}
This restricted form of the Einstein equations \eqref{Einstein1}, \eqref{Einstein2a}, and \eqref{Einstein3b} is satisfied by both classes of solutions---the Halilsoy and Chandrasekhar solutions. At this point our analysis splits into two branches. The first one with $Y(\rho)=0$ will guide us to the Halilsoy solutions. 

\subsubsection{Halilsoy solutions}

The left-hand side of Eq.\ \eqref{Einstein3b} equated to zero corresponds to the Bessel equation. In order to see this, we introduce the dimensionless coordinate $x=\rho/\lambda$ and an auxiliary function $\hat\omega=\rho\,\omega$. For $Y(\rho)=0$, the additional simplifying assumption 
\begin{equation}\label{Einstein3sim}
\dot\psi\dot\omega-\psi'\omega'=0
\end{equation}
reduces Eq.\ \eqref{Einstein3b} to the Bessel equation of order $1$
\begin{equation}\label{Einstein4b}
	\frac{d^2\hat\omega}{dx^2}+\frac{1}{x}\frac{d\hat\omega}{dx}+\left(1-\frac{1}{x^2}\right)\hat\omega=0\;.
\end{equation}
The regularity at the axis $\rho=x=0$ implies that $\omega(t,0)=0$ so $\hat\omega=o(1/x)$  at $x=0$. Therefore, we exclude from solutions the Bessel functions of the second kind and obtain
\begin{equation}
	\omega=C(t)\,\rho\, J_1\;,
\end{equation}
where with \mbox{$J_i=J_i(\rho/\lambda)$} we denote the Bessel functions of the first kind and order $i$ and where \mbox{$\ddot C(t)+1/\lambda^2C(t)=0$.} Without a loss of generality we set
\begin{equation}\label{wH}
	\omega=D\,\rho\, J_1\sin{( t/\lambda)}\;,
\end{equation}
where $D$ is a constant. The phase of a sine function is irrelevant because it can be absorbed by the choice of the time coordinate. Substituting \eqref{wH} into Eq.\ \eqref{Einstein3sim} we obtain
\begin{equation}\label{Einstein5}
	\dot\psi J_1\cos{( t/\lambda)}=\psi' J_0\sin{( t/\lambda)}\;.
\end{equation}
The general solution to Eq.\ \eqref{Einstein5} is found by the method of characteristics
\begin{equation}
	\psi=\hat\psi\circ y\;,
\end{equation}
where $y(t,\rho)$ is a function of the form
\begin{equation}\label{y}
	y(t,\rho)=J_0\cos{( t/\lambda)}\;.
\end{equation}
Utilizing this fact, we can rewrite Eq.\ \eqref{Einstein2a} as
\begin{equation}\label{Einstein6}
	\frac{d^2\hat\psi}{dy^2}+\frac{D^2}{2}e^{4\hat\psi}=0\;.
\end{equation}
The solution to Eq.\ \eqref{Einstein6} can be easily found and gives \mbox{$\psi=\hat\psi\circ y$}
\begin{equation}
\label{eq:psiAZ}
	e^{-2\psi}=\frac{|D|}{2\sqrt{c_1}} \cosh\left[2\sqrt{c_1}\,(y(t,\rho)-c_2)\right]\;,
\end{equation}
where $c_1$ and $c_2$ are integration constants of Eq.\ \eqref{Einstein6}. Using Eqs.\ \eqref{Einstein1} we obtain $\gamma$ via quadratures
\begin{equation}\label{gamma}
\gamma=\frac{c_1}{2}\left[ \left(\frac{\rho}{\lambda}\right)^2 (J_0^2+J_1^2)- 2 \frac{\rho}{\lambda} J_0 J_1 \cos^2(t/\lambda) \right].
\end{equation}
The solution to the Einstein equations presented above depends on four real parameters: $\lambda$, $c_1$, $c_2$, $D$. In what follows, we show that the number of parameters can be reduced to $3$ using a coordinate transformation.

Without a loss of generality we restrict our analysis to $D\leq 0$, where nonpositive $D$ is chosen for consistency with the Halilsoy paper \cite{Halilsoy:1988}. If $D=0$ and $c_1>0$ then the metric \eqref{metricH} becomes degenerate. However, if both parameters approach zero simultaneously $c_1\sim D^2$, then the Minkowski metric in nonstandard coordinates is recovered in the limit. Therefore, it is convenient to reparametrize both constants $c_1=A^2/4$, $D=-A B$, where $A\geq 0$ and $B \geq 0$ are new constants. The coefficients in reparametrization are conventional and were chosen again for consistency with the Halilsoy article. 

It is illustrative to rewrite Eq.\ \eqref{eq:psiAZ} in terms of the new parameters
\begin{equation}\label{psiAZ2}
	e^{-2\psi}=B \cosh\left[A\,(y-c_2)\right]\;.
\end{equation}
In the new parametrization $A=0$ corresponds to the flat metric. Although $B=0$ implies \mbox{$\omega=0$,} which in principle should correspond to the Einstein--Rosen class, the metric becomes degenerate in this limit. It turns out that dependence of metric functions on $B$ is merely a linear scaling of coordinates, and can be trivially transformed away by
\begin{equation}\label{change}
\begin{split}
	\lambda&\rightarrow \kappa\lambda\;,\\
	t&\rightarrow \kappa t\;,\\
	\rho&\rightarrow \kappa\rho\;,\\
	z&\rightarrow z/\kappa\;,
\end{split}
\end{equation}
where $\kappa=1/\sqrt{B}$. After this transformation parameter $B$ does not appear in any component of the metric. The polarization of the standing wave depends only on parameter $c_2$. 

The form of the metric given by Halilsoy can be recovered in two additional steps. First, we redefine parameter $c_2$ as follows
\begin{equation}\label{c2redef2}
	c_2=\frac{1}{A}\log\left[\operatorname{coth}\left(\alpha/2\right)\right]\;,
\end{equation}
where $\alpha\rightarrow 0$ for $A\neq 0$ corresponds to $c_2\rightarrow\infty$.
Next, we rescale coordinates using transformation of the form \eqref{change} with $\kappa=\sqrt{\sinh{\alpha}}$. 
These two steps lead to the original form of the metric given by Halilsoy \cite{Halilsoy:1988} which is parametrized by $\lambda$, $A$, $\alpha$
\begin{equation}\label{solH2}
\begin{split}
	e^{-2\psi}=&e^{A J_0 \cos{(t/\lambda)}}\sinh^2\frac{\alpha}{2}+e^{-AJ_0 \cos{(t/\lambda)}}\cosh^2\frac{\alpha}{2}\;,\\
	\omega=&-(A\sinh\alpha)\,\rho\, J_1\sin{( t/\lambda)}\;,\\
	\gamma=&\frac{1}{8}A^2\left[\left(\frac{\rho}{\lambda}\right)^2(J_0^2+J_1^2)-2\frac{\rho}{\lambda} J_0 J_1\cos^2(t/\lambda)\right]\;.
\end{split}
\end{equation}
The limit $\alpha=0$ implies $\omega=0$ and corresponds to Einstein--Rosen waves while $A=0$ corresponds to the flat metric.

Halilsoy derived his solution \eqref{solH2} in a different way than the one presented above. He used the Ernst equation and applied a generating technique to the $\omega=0$ seed to obtain the exact solutions with $\omega\neq0$ \cite{Halilsoy:1988}.

For $\alpha=0$ the solution \eqref{solH2} has only plus polarization mode and represents standing Einstein--Rosen gravitational waves investigated originally by Stephani \cite{Stephani:2003}. This solution was also studied in the context of chaos \cite{szybkanaqvi}. Its backreaction in the high-frequency limit was investigated in the article \cite{ers} where in addition a scalar field was present.

\subsubsection{Chandrasekhar solutions}

Another class of the solutions to the vacuum Einstein equations \eqref{Einstein1}--\eqref{Einstein2b} were derived by Chandrasekhar \cite{chandra}, again with the help of the Ernst equation. He restricted his analysis to the solutions with vanishing C-energy flux, namely with $\dot\gamma=0$. If $\dot\gamma=\omega=0$, then the system \eqref{Einstein1}--\eqref{Einstein2b} has only trivial solutions, hence, the Chandrasekhar solutions always have nontrivial polarization.

Our starting point is Eqs.\  \eqref{Einstein2a}, \eqref{Einstein3b} which were obtained from the original system under the assumption \eqref{harmonicO}. In contrast to previous analysis, which guided us to the Halilsoy solution, we do not impose $Y(\rho)=0$ and we do not adopt the simplifying assumption \eqref{Einstein3sim}. Instead, constant C-energy, $\dot\gamma=0$, implies, by the first equation in Eq.\ \eqref{Einstein1}, that
\begin{equation}\label{Einstein2c}
	0=2\rho\dot\psi\psi'+\frac{e^{4\psi}}{2\rho}\dot\omega\omega'\;.
\end{equation}
In order to simplify the equation above we introduce a new function $\hat\Psi$ such that $$\psi=-\frac{1}{2}\ln\hat\Psi\;.$$
Equation \eqref{Einstein2c} takes the form
\begin{equation}\label{Einstein2d}
	0=\rho^2\dot{\hat\Psi}\hat\Psi'+\dot\omega\omega'\;.
\end{equation}
Since $\omega$ satisfies the harmonic oscillator equation \eqref{harmonicO}, we can write
\begin{equation}\label{hoomega}
\omega=\lambda\left(X\cos(t/\lambda)+Y\right)\;,
\end{equation}
where $X$ and $Y$ are arbitrary dimensionless functions of $\rho$ and where the phase of $\omega$ in $t$ was fixed without a loss of generality. In analogy to the assumption \eqref{harmonicO}, we further simplify the problem by demanding that $\hat\Psi$ depends on $t$ in the simplest nontrivial way, namely that it solves the harmonic oscillator equation in $t$ with the same phase as $\omega$. We have
\begin{equation}\label{hoP}
\hat\Psi=Z\cos(t/\lambda)+W\;,
\end{equation}
where $Z$ and $W$ are arbitrary dimensionless functions of $\rho$. We rewrite Eqs.\ \eqref{Einstein2a}, \eqref{Einstein2b}, and \eqref{Einstein2d} in terms of the functions $X$, $Y$, $Z$, $W$. These equations are too large to be usefully presented here. They depend on time only through terms that can be presented as powers of $\cos(t/\lambda)$. Since these equations hold for any $t$, then all coefficients in front of such terms have to vanish. Equation \eqref{Einstein2a} gives two equations of the first order and each of the remaining two equations reduce to three equations of the second order, so we have eight ordinary differential equations in total for $X$, $Y$, $Z$, $W$, and their derivatives. We determine $X''$, $Y''$, $Z''$, and $W''$ using four equations and substitute the second derivatives into the remaining four equations. The resulting system consists of four equations of the first order. We use two of them to calculate $X'$, $Y'$
\begin{equation}\label{XpYp}
\begin{split}
	X'=&-\frac{\rho^2}{\lambda^2}\frac{Z}{X}Z'\;,\\
	Y'=&-\frac{\rho^2}{\lambda^2}\frac{Z}{X}W'\;.
\end{split}
\end{equation}
We substitute $X'$ and $Y'$ to the remaining equations. One of the equations is 
\begin{equation}\label{xyzw1}
	\begin{split}
	&\left(\lambda^2 X^2-\rho^2 Z^2\right)\cdot\\
	&\left(X^2(Z^2-W^2)+\rho^2(ZW'-WZ')^2\right)=0\;.
	\end{split}
\end{equation}
The second equation is equivalent to the vanishing of the factor in the second parenthesis in Eq.\ \eqref{xyzw1}, so in fact we are left with a single equation

\begin{equation}
\label{xyzw1a}
X^2(Z^2-W^2)+\rho^2(ZW'-WZ')^2=0\;.
\end{equation}
The first term is necessarily nonpositive. Therefore, we introduce auxiliary functions $R=R(\rho)$ and $P=P(\rho)$
\begin{equation}
	W=R\cosh P\;,\quad Z=R\sinh P\;,
\end{equation}
such that $Z^2-W^2=-R^2$. 
Equation \eqref{xyzw1a} can be rewritten as
\begin{equation}\label{xyzw2}
	P'=\pm \frac{1}{\rho}\frac{X}{R}\;,
\end{equation}
where the $\pm$ sign is equivalent up to reflections $\phi\rightarrow -\phi$ (we chose $+$). In order to match the notation of Chandrasekhar \cite{chandra} we introduce one more auxiliary function $F=F(\rho)$ such that $|F|<1$
\begin{equation}\label{Pdef}
	P=2\operatorname{arctanh}(-F)\;.
\end{equation}
\forceeqnum{Pdef}
So far we have only redefined functions and conducted algebraic manipulations treating $X$, $Y$, $W$, $Z$ and their derivatives as independent variables. The differential relations between these functions and their derivatives lead to $R=const$ (in fact regularity at $\rho=0$ implies $R=1$) and to the second order ordinary differential equation for $F(\rho)$.

The functions $X$, $Y'$, $Z$, $W$ can be presented in terms of $F$ and its derivatives
\begin{equation}\label{xyzw3}
\begin{split}
	X=&2\rho\frac{F'}{1-F^2}\;,\\
	Y'=&4\frac{\rho}{\lambda^2}\frac{F^2}{(1-F^2)^2}\;,\\
	W=&\frac{1+F^2}{1-F^2}\;,\\
	Z=&-\frac{2F}{1-F^2}\;.
\end{split}
\end{equation}
The equation for the function $F$ can be obtained directly from Eq.\ \eqref{XpYp}. We observe that
\begin{equation}\label{F2a}
\frac{X'}{Y'}=\frac{Z'}{W'}\;.
\end{equation}
In terms of $F$ the equation above reads
\begin{equation}\label{F}
	F(1+F^2)/\lambda^2+\frac{1-F^2}{\rho}(\rho F')'+2F (F')^2=0\;.
\end{equation}
The regularity of solutions at $\rho=0$ implies the boundary conditions $F(0)=F_0$, $F'(0)=0$, where $0<F_0<1$ is a constant. 

These solutions correspond to Chandrasekhar solution. Changing the notation to match Chandrasekhar's formulas \cite{chandra}, we have
\begin{equation}\label{Psi}
	\Psi=\frac{1}{\hat\Psi}=\frac{1-F^2}{1+F^2-2F\cos(t/\lambda)}\;.
\end{equation}
Similarly, we introduce a second function $\Omega$ (for consistency with the paper \cite{NikielSzybka:2025}), which corresponds to $-q_2$ in the Chandrasekhar notation \cite{chandra})
\begin{equation}\label{Omega}
	\Omega/\lambda=\omega/\lambda=\frac{2\rho F'}{1-F^2}\cos{(t/\lambda)}+\frac{4}{\lambda^2}\int_0^\rho\frac{\hat\rho F^2}{(1-F^2)^2}d\hat\rho\;,
\end{equation}
where $F=F(\rho)$. With this choice of $\Psi$ and $\Omega$ we set $\psi=\ln\sqrt{\Psi}$, $\omega=\Omega$, $\nu=\gamma-\ln\sqrt{\Psi}$, where $\nu=\nu(t,\rho)$ is a new metric function to match the Chandrasekhar notation. The Einstein equations \eqref{Einstein1} reduce to 
\begin{equation}\label{nu}
\begin{split}
	(\nu+\ln\sqrt\Psi)^{\Ldot}=&0\;,\\
	(\nu+\ln\sqrt\Psi)'=&\frac{\rho}{(1-F^2)^2}(F^2/\lambda^2+ (F')^2)\;.
\end{split}
\end{equation}
\forceeqnum{nu}
The essential part of the Einstein equations \eqref{Einstein2a}, \eqref{Einstein2b} has the form of ordinary differential equation for $F(\rho)$ and is given by Eq.\ \eqref{F}.

The line element \eqref{metricH} in terms of $\Psi$, $\nu$, and $\Omega$ reads
\begin{equation}\label{metricCh}
	\hat{g}=e^{2\nu}\left(-dt^2+d\rho^2\right)+\frac{\rho^2}{\Psi}d\phi^2+\Psi(dz+\Omega d\phi)^2\;,
\end{equation}
where the coordinates have a standard range.

\section{High-frequency limit}\label{highfreq}

It was shown in the paper \cite{NikielSzybka:2025} that the Halilsoy and Chandrasekhar solutions differ in the linear approximation. It is therefore interesting to investigate whether they share the same high-frequency limit, i.e., whether they can be represented by the same effective spacetime as frequency of standing gravitational waves tends to infinity \mbox{($\lambda\rightarrow 0$).} In the next section, we derive the effective energy-momentum tensors that represent ``averaged'' energy-momentum content for both classes of spacetimes. To conduct the analysis, we identify specific one-parameter families of solutions in each class that yield nontrivial backreaction effects in the high-frequency limit. The existence of such families is not guaranteed {\it a priori} and is interesting in its own right \cite{Huneau:2024lrb}.

At the linear order, the essential difference between the Halilsoy and Chandrasekhar solutions reduces to different polarizations of the standing waves, so one might expect the polarization not to affect the effective energy-momentum content. In the following sections, we show that this is indeed the case.

In this paper, we consider only vacuum spacetimes. The high-frequency limit, also called the weak limit, can be computed for vacuum spacetimes using Burnett's method \cite{burnett}. Here we follow the more general Green--Wald formulation \cite{greenwald} which is also applicable to nonvacuum solutions.

We define, as in the paper \cite{greenwald}
$$h_{\alpha\beta}(\lambda,x)=g_{\alpha\beta}(\lambda,x)-g^{(0)}_{\alpha\beta}(x)\;,$$
where $g^{(0)}_{\alpha\beta}$ is a smooth background metric. This background metric is approached by a one-parameter family of metrics $g_{\alpha\beta}(\lambda,x)$ as $\lambda\rightarrow 0$. A nontrivial backreaction effect arises for families of solutions in which covariant derivatives $h_{\alpha\beta}(\lambda,x)$ (relative to $g^{(0)}_{\alpha\beta}$) fail to converge pointwise as $\lambda\rightarrow 0$. We assume that these derivatives vanish in the weak sense as $\lambda\rightarrow 0$. We place these considerations on a rigorous footing using the notion of weak limit defined in Ref.~\cite{greenwald}. 

Letting $A_{\alpha_1\ldots\alpha_n}(\lambda,x)$ be a family of tensor fields defined for $\lambda>0$, we say that $A_{\alpha_1\ldots\alpha_n}(\lambda,x)$ converges weakly to $B_{\alpha_1\ldots\alpha_n}(x)$ as $\lambda\rightarrow 0$ if for all smooth $f^{\alpha_1\ldots\alpha_n}(x)$ of compact support, we have
\begin{equation}
\begin{split}
	&\lim_{\lambda\rightarrow 0}\int f^{\alpha_1\ldots\alpha_n}(x)A_{\alpha_1\ldots\alpha_n}(\lambda,x)\sqrt{|g^{(0)}|}d^n x\\
	&=\int f^{\alpha_1\ldots\alpha_n}(x)B_{\alpha_1\ldots\alpha_n}(x)\sqrt{|g^{(0)}|}d^n x\;.
\end{split}
\end{equation}
The weak limit can be interpreted as a form of ``spacetime averaging.'' Adopting the convention of Green and Wald, we denote it with $\mathop{\mathrm{w\text{-}lim}}$. 

It is not immediately clear how to translate this definition into an operational procedure. However, in the case studied here, the following elementary observation proves very useful
\begin{equation}
	\mathop{\mathrm{w\text{-}lim}}_{\lambda\rightarrow0}\cos^2(x/\lambda)=\mathop{\mathrm{w\text{-}lim}}_{\lambda\rightarrow0}\frac{1}{2}\left[1+\cos(2x/\lambda)\right]=\frac{1}{2}\;.
\end{equation}
Similar trigonometric manipulations allow us to complete all necessary calculations of weak limits.

Burnett conjectures existence of one-parameter solutions $g_{\alpha\beta}(\lambda,x)$ to the vacuum Einstein equations satisfying the following conditions \cite{greenwald}.
\begin{enumerate}[(i)]
\item For each value of the parameter $\lambda>0$ the vacuum Einstein equations hold
\begin{equation}
	G_{\alpha\beta}(\lambda)=0\;.
\end{equation}
\item There exists a smooth positive function $C_1(x)$ such that
\begin{equation}
|h_{\alpha\beta}(\lambda,x)|\leq\lambda C_1(x)\;.
\end{equation}
\item There exists a smooth positive function $C_2(x)$ such that
\begin{equation}
|\nabla_\mu h_{\alpha\beta}(\lambda,x)|\leq C_2(x)\;.
\end{equation}
\item There exists a smooth tensor field $\mu_{\alpha\beta\gamma\delta\mu\nu}$ such that
\begin{equation}
\mathop{\mathrm{w\text{-}lim}}_{\lambda\rightarrow 0}\nabla_\alpha h_{\gamma\delta}(\lambda)\nabla_\beta h_{\mu\nu}(\lambda)=\mu_{\alpha\beta\gamma\delta\mu\nu}\;.
\end{equation}
\end{enumerate}

The norm $|.|$ used above can be taken relative to any Riemannian metric (we use the Euclidean norm). The nonvacuum extension of the Burnett's method presented by Green and Wald \cite{greenwald} modifies the condition (i) to the nonvacuum case $G_{\alpha\beta}(\lambda)=8\pi T(\lambda)$, where $T(\lambda)$ is the energy-momentum tensor satisfying the weak energy condition.

\section{Backreaction}\label{BB}

The backreaction effect is represented by the effective energy-momentum tensor which is given by 
\begin{equation}\label{t0}
t^{(0)}_{\alpha\beta}=\frac{1}{8\pi}G_{\alpha\beta}(g^0)\;.
\end{equation}

The properties of $t_{\alpha\beta}$ are {\it a priori} unknown. Burnett showed that $t_{\alpha\beta}$ can be also calculated in terms of the tensor $\mu_{\alpha\beta\gamma\delta\mu\nu}$ (defined above). The properties of this tensor allow one to prove that $t_{\alpha\beta}$ is traceless, which is important for cosmological backreaction in a nonvacuum extension of the Burnett's method presented by Green and Wald \cite{greenwald}.

We have \cite{burnett}
\begin{equation}\label{t0fmu}
\begin{split}
8\pi\,t_{\alpha\beta}^{(0)}=\;\,&\frac{1}{8}\Bigl(-\mu\indices{^\gamma_\gamma^{\delta\mu}_{\delta\mu}}-\mu\indices{^\gamma_\gamma^\delta_\delta^\mu_\mu}+2\,\mu\indices{^{\gamma\delta}_\gamma^\mu_{\delta\mu}}\Bigr)
	g_{\alpha\beta}^{(0)}\\&+\frac{1}{2}\,\mu\indices{^{\gamma\delta}_{\alpha\gamma\beta\delta}}-\frac{1}{2}\,\mu\indices{^\gamma_{\gamma\alpha}^\delta_{\beta\delta}}
	+\frac{1}{4}\,\mu\indices{_{\alpha\beta}^{\gamma\delta}_{\gamma\delta}}\\&-\frac{1}{2}\,\mu\indices{^\gamma_{(\alpha\beta)\gamma}^\delta_{\delta}}
+\frac{3}{4}\,\mu\indices{^\gamma_{\gamma\alpha\beta}^\delta_{\delta}}-\frac{1}{2}\,\mu\indices{^{\gamma\delta}_{\alpha\beta\gamma\delta}}\;.
\end{split}
\end{equation}
The tensor $\mu_{\alpha\beta\gamma\delta\mu\nu}$ is symmetric in pairs of indices $\mu_{\alpha\beta\gamma\delta\mu\nu}=\mu_{(\alpha\beta)(\gamma\delta)(\mu\nu)}$. In addition, $\mu_{\alpha\beta\gamma\delta\mu\nu}=\mu_{\alpha\beta\mu\nu\gamma\delta}$. These symmetries imply that out of $4^6=4096$ components only $10\cdot\frac{10^2-10}{2}=450$ are independent. According to Burnett \cite{burnett}, the effective energy-momentum tensor $t^{(0)}_{\alpha\beta}$ can also be written in terms of the tensor 
\begin{equation}\label{alpha}
\alpha_{\alpha\beta\gamma\delta\mu\nu}=\mu_{[\gamma|[\alpha\beta]|\delta]\mu\nu}\;.
\end{equation}
(To preserve consistency with other articles, we abuse notation and use the same symbol $\alpha$ as a name of the tensor, the index and the polarization parameter in the Halilsoy solution.)
We have \cite{burnett}
\begin{equation}\label{t0fa}
	8\pi\,t_{\alpha\beta}^{(0)}=\alpha\indices{_{\alpha}^{\gamma}_{\beta}^{\delta}_{\gamma\delta}}\;,
\end{equation}
where
\begin{equation}\label{talpha}
	\alpha\indices{_{\alpha}^{\gamma}_{\beta\gamma\mu\nu}}=0\;.
\end{equation}

\subsection{Halilsoy solutions}

The construction of a one-parameter family of solutions satisfying in a nontrivial way the Burnett's conjecture (see Sec.\ \ref{highfreq}) is straightforward. In analogy to the article \cite{ers}, it is sufficient to set the constant \mbox{$A=2\beta \sqrt{\lambda}$}, where $\beta>0$ is a new constant and where $\lambda$ is the main parameter. The factor $2$ was chosen for a consistency with the article \cite{ers}.

The inspection of the metric functions \eqref{solH2} shows that in the limit $\lambda\rightarrow 0$ we have $\Psi^{(0)}=-\ln\sqrt{\cosh\alpha}$, $\omega^{(0)}=0$. In order to calculate $\gamma^{(0)}$ we assume $\rho>\rho_0>0$ and use the expansion of the Bessel functions for $\lambda\rightarrow 0$ to find $\gamma^{(0)}=\beta^2\rho/\pi$.
 This gives the background metric for $\rho>\rho_0$
\begin{equation}\label{g0H}
	\begin{split}
		g^{(0)}&=\cosh\alpha\Big(e^\frac{2\beta^2\rho}{\pi}\left(-dt^2+d\rho^2\right)+\rho^2d\phi^2\\&+\sech^2\!{\alpha}\,dz^2\Big)\;,
	\end{split}
\end{equation}
which can be analytically extended to $\rho>0$.
This metric depends on the parameter $\alpha$ in a trivial way---it can be removed by a coordinate transformation as was done in the paper \cite{NikielSzybka:2025}. This is a null dust spacetime discovered before by Morgan \cite{morgan}. It has a curvature singularity at $\rho=0$.

The condition (i) is satisfied because the Halilsoy solutions are exact solutions to the vacuum Einstein equations. It is easy to see that the conditions (ii) and (iii) are also satisfied. We will show by a direct calculation (with the help of Wolfram {\it Mathematica}) that the condition (iv) holds. In fact, the explicit form of the tensor $\mu_{\alpha\beta\gamma\delta\mu\nu}$ is needed only to validate the formalism. The backreaction effect can be calculated directly from the Einstein tensor using the formula \eqref{t0}.

The effective energy-momentum tensor is given by
\begin{equation}\label{t0res}
	t^{(0)}=\frac{G(g^{(0)})}{8\pi}=\frac{\beta^2}{8\pi^2\rho}\left(dt^2+d\rho^2\right)\;,
\end{equation}
which coincides with the result in the article \cite{ers} for plus polarized solutions without a scalar field. The nontrivial polarization of standing waves does not contribute to the backreaction effect. The effective energy-momentum tensor is traceless, as expected, and of the Segre type $[(11)1,1]$ which aligns with the ideas presented in the article \cite{Szybka:2019}.

In order to check the consistency of Burnett's method and its extension to the Green--Wald framework, we have to calculate the tensor $\mu_{\alpha\beta\gamma\delta\mu\nu}$. We have
\begin{equation}
\begin{split}
	&\mu_{tttttt}=-\mu_{tttt\rho\rho}=\mu_{tt\rho\rho\rho\rho}=\mu_{\rho\rho tttt}=-\mu_{\rho\rho tt\rho\rho}=\mu_{\rho\rho\rho\rho\rho\rho}\\
        &=e^{4\beta^2\rho/\pi}\frac{2}{\pi}\beta^2\rho^{-1}\left[1+\frac{1}{4\pi}\beta^2\rho\,(1+\cosh{2\alpha})\right]\,,\\
	&\mu_{tttt\phi\phi}=-\mu_{tt\rho\rho\phi\phi}=\mu_{\rho\rho tt\phi\phi}\\
	&=-\mu_{\rho\rho\rho\rho\phi\phi}=-\frac{2}{\pi}\beta^2\rho\, e^{2\beta^2\rho/\pi}\,,\\
        &\mu_{ttttzz}=-\mu_{tt\rho\rho zz}=\mu_{\rho\rho ttzz}\\
	&=-\mu_{\rho\rho\rho\rho zz}=\frac{2}{\pi}\beta^2\rho^{-1}e^{2\beta^2\rho/\pi}\sech^2\!{\alpha}\,,\\
&\mu_{tt\phi\phi\phi\phi}=\mu_{\rho\rho\phi\phi\phi\phi}=\frac{2}{\pi}\beta^2\rho^3\,,\\
	&\mu_{tt\phi\phi zz}=\mu_{\rho\rho\phi\phi zz}=-\frac{2}{\pi}\beta^2\rho\sech^2\!{\alpha}\,,\\
        &\mu_{ttzzzz}=\mu_{\rho\rho zzzz}=\frac{2}{\pi}\beta^2\rho^{-1}\sech^4\!{\alpha}\,,\\
	&\mu_{tt\phi z\phi z}=\mu_{\rho\rho\phi z\phi z}=\frac{2}{\pi}\beta^2\rho\tanh^2\!{\alpha}\,,\\
	&\mu_{t\rho tt \phi z}=-\mu_{t\rho\rho\rho\phi z}=\frac{2}{\pi}\beta^2 e^{2\beta^2\rho/\pi}\tanh{\alpha}\,,\\
	&\mu_{t\rho\phi\phi\phi z}=-\frac{2}{\pi}\beta^2\rho^2\tanh{\alpha}\,,\\
	&\mu_{t\rho\phi zzz}=\frac{2}{\pi}\beta^2\sech^2\!\alpha\tanh{\alpha}\,.\\
\end{split}
\end{equation}
The remaining components follow from symmetries or vanish. The tensor $\mu_{\alpha\beta\gamma\delta\mu\nu}$ for $\alpha=0$ coincides with the one presented in the article \cite{ers} for the plus polarized solutions without a scalar field. We have calculated the effective energy-momentum tensor using Eq.\ \eqref{t0fmu} and it coincides with one calculated directly from the Einstein tensor \eqref{t0}, \eqref{t0res}. Next, we have calculated the tensor $\alpha_{\alpha\beta\gamma\delta\mu\nu}$ using Eq.\ \eqref{alpha} and we verified that the condition \eqref{talpha} holds. Finally, we have determined the effective energy-momentum tensor using Eq.\ \eqref{t0fa}. All three results [Eqs.\ \eqref{t0res}, \eqref{t0fmu}, and \eqref{t0fa}] are the same, which confirms the consistency of the Burnett, Green--Wald approach.

\subsection{Chandrasekhar solutions}

In this section we calculate the backreaction effect for the Chandrasekhar solutions. In contrast to the Halilsoy solutions, which were studied in the previous section, the Chandrasekhar solutions are not exact. The essential part of the Einstein equations is given by Eq.\ \eqref{F}. The exact solutions of this equation are not known, therefore it has to be solved numerically. In order to evade this difficulty, we construct the one-parameter family of spacetimes $[M,g(\lambda)]$ which for small but nonzero $\lambda$ approximates to the one-parameter numerical family in the Chandrasekhar class given by Eq.\ \eqref{metricCh} and the metric functions $\Psi$, $\nu$, $\Omega$ defined above this equation. This family of spacetimes $[M,g(\lambda)]$ is given by exact metric functions and satisfies the conditions (ii), (iii), (iv) of the Burnett's conjecture presented in Sec.\ \ref{highfreq}. The condition (i) is satisfied for the nonvacuum Einstein equations with some $T(\lambda)=G[g(\lambda)]/(8\pi)$. Since $g(\lambda)$ approximates vacuum solutions for small $\lambda$, $T(\lambda)$ remains small for small but nonzero $\lambda$. The tensor $T(\lambda)$ has a very complicated functional form, so we did not verify whether it satisfies the weak energy condition. However, we calculated the weak limit of $T(\lambda)$ and we found that it vanishes. Therefore, it does not contribute to the backreaction effect and the analysis presented in this section differs trivially from the vacuum case.

Our starting point is the nonlinear equation \eqref{F} for the function $F(\rho)$, which defines the remaining metric functions. The Minkowski spacetime corresponds to $F=0$. We set $F(0)=\beta\sqrt{\lambda/2}$ to satisfy the condition (ii) from Sec.\ \ref{highfreq} ($\beta$ is a constant such that $0<\beta\sqrt{\lambda/2}<1$). It will become clear later why we have chosen the square root of $\lambda$ instead of $\lambda$. Since we are looking for high frequency periodic or quasiperiodic solutions, we substitute $F(\rho)=\beta\sqrt{\lambda/2} f[x(\rho)]$ into \eqref{F}, where $x=\rho/\lambda$ for $\lambda>0$. The initial condition takes the form $f(0)=1$ and the regularity implies $f'(0)=0$. The equation for $f(x)$ reads (primes denote derivatives in $x$)
\begin{equation}\label{f}
	f''+f'/x+f-\sqrt\lambda\beta(f''+f'/x)f+\lambda\beta^2(2f'^2+f^2)f=0\;.
\end{equation}
For small $\lambda$ this equation can be approximated by the Bessel equation. The initial condition $f(0)=1$ implies 
\begin{equation}
	F(\rho)=\beta\sqrt{\lambda/2} f(\rho/\lambda)=\beta\sqrt{\lambda/2} J_0(\rho/\lambda)\;,
\end{equation}
which is consistent with the regularity condition \mbox{$f'(0)=0$.}

For a moment we restrict our analysis to $\rho>\rho_0>0$. For small $\lambda$ the function $F(\rho)$ can be approximated by 
\begin{equation}
	F(\rho)\approx\frac{\beta\lambda}{\sqrt{\pi\rho}} \cos(\pi/4-\rho/\lambda)\;,
\end{equation}
which gives
\begin{equation}
	F'(\rho)\approx\frac{\beta}{\sqrt{\pi\rho}} \sin(\pi/4-\rho/\lambda)\;,
\end{equation}
\begin{equation}
	F'(\rho)\approx\frac{\beta}{\sqrt{\pi\rho}} \sin(\pi/4-\rho/\lambda)\; -\frac{\beta \lambda}{\sqrt{\pi}\rho^{3/2}} \cos{(\pi/4-\rho/\lambda)}\;.
\end{equation}
We use these approximations to calculate the metric functions $\Psi$, $\Omega$, $\nu$ in the limit $\lambda\rightarrow 0$. Equation \eqref{Psi} implies $\Psi^{(0)}=1$ and Eq.\ \eqref{Omega} gives $\Omega^{(0)}=0$. The equations for the metric function $\nu$ can be approximated by
\begin{equation}
\begin{split}
	\nu^{\Ldot}=&0\;,\\
\nu'=&\frac{\beta^2}{\pi}\;,
\end{split}
\end{equation}
hence $\nu^{(0)}=\beta^2\rho/\pi$ (the additive constant can be removed by the change of coordinates). The background metric for $\rho>\rho_0$ reads 
\begin{equation}\label{g0Ch}
        g^{(0)}=e^\frac{2\beta^2\rho}{\pi}\left(-dt^2+d\rho^2\right)+\rho^2d\phi^2+dz^2\;,
\end{equation}
and can be analytically extended to the whole region $\rho>0$. It is, again, the Morgan solution \cite{morgan}. This metric differs from the metric \eqref{g0H} only in the choice of coordinates. Therefore, the backreaction effect is the same as for the Halilsoy solution and is given by the effective energy-momentum tensor \eqref{t0res}. 

We also calculate components of tensor $\mu_{\alpha\beta\gamma\delta\mu\nu}$ for the Chandrasekhar solution. Its nonzero components are
\begin{equation}
\begin{split}
	&\mu_{tttttt}=-\mu_{tttt \rho\rho}=\mu_{tt\rho\rho\rho\rho}=\mu_{\rho\rho tttt}=-\mu_{\rho\rho tt \rho\rho}=\\
	&=\mu_{\rho\rho\rho\rho\rho\rho}=\frac{e^{4\beta^2\rho/\pi}\beta^2}{\pi\rho}\,,\\
	&\mu_{tttt\phi\phi}=-\mu_{tt\rho\rho\phi\phi}=\mu_{\rho\rho tt \phi\phi}=-\mu_{\rho\rho\rho\rho\phi\phi}=-\frac{e^{2\beta^2\rho/\pi}\beta^2\rho}{\pi}\,,\\
	&\mu_{ttttzz}=-\mu_{tt \rho\rho zz}=\mu_{\rho\rho tt zz}=-\mu_{\rho\rho\rho\rho zz}=\frac{e^{2\beta^2\rho/\pi}\beta^2}{\pi\rho}\,,\\
	&\mu_{tt\phi\phi\phi\phi}=\mu_{\rho\rho\phi\phi\phi\phi}=\frac{\beta^2\rho^3}{\pi}\,,\\
	&\mu_{tt\phi \phi zz}=-\mu_{tt \phi z \phi z}=\mu_{\rho\rho\phi \phi zz}=-\mu_{\rho\rho\phi z \phi z}=-\frac{\beta^2 \rho}{\pi}\,,\\
        &\mu_{ttzzzz}=\mu_{\rho\rho zzzz}=\frac{\beta^2}{\pi\rho}\;,\\
\end{split}
\end{equation}
where the remaining components follow from symmetries of $\mu_{\alpha\beta\gamma\delta\mu\nu}$. These components do not coincide with corresponding components of this tensor in the case of the Halilsoy solution, not even for some specific choice of polarization parameter $\alpha$ of the Halilsoy solution. Nonetheless, the resulting effective energy-momentum tensor, calculated using Eq.\ \eqref{t0fmu}, is still the same as the one given by Eq.\ \eqref{t0res}. We verified that the condition \eqref{talpha} holds and that the formulas \eqref{t0res}, \eqref{t0fmu}, \eqref{t0fa} lead to the same results. In summary, the analysis of this section further confirms the consistency of the Burnett, Green--Wald approach.

\section{Properties of the effective metric}

The high-frequency limit of both solutions---the Halilsoy solution \eqref{metricH}, \eqref{solH2} and the Chandrasekhar solution \eqref{F}--\eqref{metricCh}---given by Eq.\ \eqref{g0Ch}, corresponds to the Morgan solution, which is probably the simplest nontrivial metric that satisfies the Einstein equations with a physical matter content \cite{morgan}. The regularity of components of the metric is misleading here. The blow up of the Kretschmann scalar leads to a curvature singularity at \mbox{$\rho=0$} which is naked. This cylindrically symmetric metric is not asymptotically flat in the radial direction. Because of that, despite its simple form, it has received little attention in the literature.  We briefly comment on its properties including the causal structure. 

The Morgan metric cannot be asymptotically flat because it is cylindrically symmetric. The behavior in the limit \mbox{$\rho\rightarrow+\infty$} for $z=const$ needs a more detailed treatment \cite{Stachel1966}. 

We choose the Newman--Penrose null tetrad $l$, $n$, $m$, $\bar{m}$ such that
\begin{equation}
\begin{split}
	l=&\frac{e^{-\frac{\beta^2\rho}{\pi}}}{\sqrt{2}}\left(\partial_t+\partial_\rho\right)\;,\\
	n=&\frac{e^{-\frac{\beta^2\rho}{\pi}}}{\sqrt{2}}\left(\partial_t-\partial_\rho\right)\;,\\
	m=&\frac{1}{\sqrt{2}}\left(\frac{1}{\rho}\partial_\phi+i\partial_z\right)\;,
\end{split}
\end{equation}
where $l\cdot n=-1$, $m\cdot\bar{m}=1$, and other scalar products vanish. Our tetrad differs from the one used in the article \cite{Stachel1966} only trivially, so their analysis of the asymptotic properties of the curvature tensor applies here.

The Weyl tensor is algebraically general (the Petrov type I). There are two nonzero Weyl scalars (we note an error in the paper \cite{Szybka:2019}) $$\Psi_0=\Psi_4=\frac{\beta^2}{2\pi\rho} e^{-\frac{2\beta^2\rho}{\pi}}\;.$$  

The Weyl scalar $\Psi_0$ does not satisfy the cylindrical peeling property presented in the paper \cite{Stachel1966}, namely $$e^{\frac{2\beta^2\rho}{\pi}}\Psi_0=\frac{\beta^2}{2\pi\rho}$$ falls off slower than than $O(1/\rho^{5/2})$ and $O(1/\rho^2)$. In contrary to that, $e^{\frac{2\beta^2\rho}{\pi}}\Psi_4$ falls off faster than $O(1/\rho^{1/2})$, hence it satisfies the peeling property.
Although the Weyl tensor, the Kretschmann scalar, and the energy density of null dust vanish at null infinities, the Morgan spacetime restricted to $z=consts$ hypersurfaces is asymptotically flat neither in the standard sense nor in the polyhomogeneous sense.

\begin{figure}[t!]
	\includegraphics[width=0.26\textwidth]{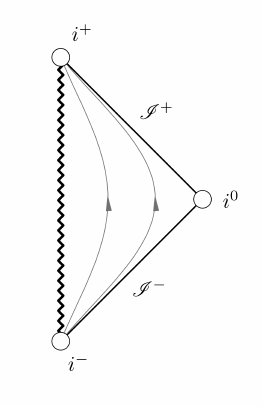}
	\caption{Penrose diagram.}
\label{Penrose}
\end{figure}

The causal structure of the spacetime can be presented in terms of the projection diagrams, introduced in the article \cite{diagrammatics}. Using methods presented there the appropriate projection diagram can be read out directly from the metric \eqref{g0Ch}. It corresponds to the Penrose diagram of the metric induced on the hypersurface $z=const$, which is presented in Fig.\ \ref{Penrose}. Nevertheless, it is instructive to construct the Penrose diagram by a direct calculation.

We introduce dimensionless null coordinates 
\begin{equation}
	u=-e^{-\frac{\beta^2}{\pi}(t-\rho)}\;,\quad v=e^{\frac{\beta^2}{\pi}(t+\rho)}\;.
\end{equation}
The inverse transformation reads
\begin{equation}
	t=\frac{\pi}{2\beta^2}\ln\left(-\frac{v}{u}\right)\;,\quad \rho=\frac{\pi}{2\beta^2}\ln\left(-u\,v\right)\;.
\end{equation}
The metric \eqref{g0Ch} can be written in the form
\begin{equation}\label{g0null}
	g^{(0)}=-\left(\frac{\pi}{\beta^2}\right)^2 du dv+\rho^2\!(u,v)d\phi^2+dz^2\;,
\end{equation}
where $u\,v<-1$ to avoid the curvature singularity (and the degeneracy of the metric). The region where $u>0$ and $v<0$ is an isometric copy of the region where $u<0$ and $v>0$, disconnected from it, and is of no interest to us. We apply the standard compactification $u=\tan{\bar u}$, $v=\tan{\bar v}$ followed by the transformation 
%\begin{samepage}
%
\begin{equation}
\bar{t}=\frac{1}{2}(\bar{u}+\bar{v})\;,\quad \bar{x}=\frac{1}{2}(\bar{v}-\bar{u})\;.
\end{equation}
This leads to
\begin{equation}\label{g0null2}
	g^{(0)}=\left(\frac{\pi}{\beta^2}\right)^2\frac{-d\bar{t}^2+d\bar{x}^2}{\cos^2\!(\bar{t}-\bar{x})\cos^2\!(\bar{t}+\bar{x})}+\rho^2(\bar{t},\bar{x})d\phi^2+dz^2\;.
\end{equation}
%
%\end{samepage}

We restrict our analysis to the hypersurface $z=const$ factoring out the symmetry in the direction $\partial_z$. It remains to establish the allowed range of coordinates in the metric \eqref{g0null2} to draw the Penrose conformal diagram. 
  
The metric \eqref{g0null2} is not well defined if the cosines in the denominator vanish. Therefore, we consider a set in $\mathbb{R}^2$ bounded by the following lines: $\bar t-\bar x\in\{\pm\pi/2\}$, $\bar t+\bar x\in\{\pm\pi/2\}$. We are interested in the region where $\rho(\bar t,\bar x)>0$ which corresponds to $u\,v<-1$. This implies $\bar v-\bar  u>\pi/2$, hence $\bar x>\pi/4$. Our region of interest is bounded by the singular line \mbox{$\bar x=\pi/4$.} It follows from our analysis that the singularity $\rho=0$ is timelike and naked. The Penrose diagram is shown in Fig.\ \ref{Penrose}. The region presented in the diagram is covered by the original coordinates $t$, $\rho$.

\section{Conclusions}

The Halilsoy \cite{Halilsoy:1988} and Chandrasekhar \cite{chandra} solutions to the vacuum Einstein equations describe cylindrically symmetric standing gravitational waves with different polarizations \cite{NikielSzybka:2025}. The Chandrasekhar solution satisfies a restrictive standing-wave condition---C-energy is time independent. By contrast, in the Halilsoy solution C-energy varies in time and is constant only on average. 

We derived both solutions directly from the Einstein equations and identified the assumptions they share and the ways in which they differ.

We proved that both solutions have the same high-frequency limit given by the Morgan null-dust solution. This limit also coincides with the high-frequency limit of Einstein--Rosen waves with a massless minimally coupled scalar field \cite{ers}---the scalar field alters only the energy density of null dust. Our analysis provides further evidence of the consistency of the Green--Wald approach to the backreaction effect \cite{greenwald} and bears on Burnett's conjecture \cite{burnett}.

We have analyzed causal and asymptotic structure of the Morgan spacetime, showing that it does not satisfy the cylindrical peeling properties. This effective cylindrical spacetime is not asymptotically flat in the radial direction. It depends on the amplitude of the waves but not on their polarization.

Some calculations were performed using Wolfram {\it Mathematica} and the {\sc xAct} package \cite{xAct}. The Penrose diagram was created using the ePiX library presented in the article \cite{Olz}.

\section*{Acknowledgments}

We gratefully acknowledge the contributions of OpenAI's GPT and Copilot for their assistance in resolving linguistic nuances.

\section*{Data availability}

The Wolfram {\it Mathematica} code is not publicly available but may be shared by the authors for verification purposes upon request.

\bibliographystyle{apsrev4-2}
\setcitestyle{authortitle}
\bibliography{report}

\begin{thebibliography}{35}%
\makeatletter
\providecommand \@ifxundefined [1]{%
 \@ifx{#1\undefined}
}%
\providecommand \@ifnum [1]{%
 \ifnum #1\expandafter \@firstoftwo
 \else \expandafter \@secondoftwo
 \fi
}%
\providecommand \@ifx [1]{%
 \ifx #1\expandafter \@firstoftwo
 \else \expandafter \@secondoftwo
 \fi
}%
\providecommand \natexlab [1]{#1}%
\providecommand \enquote  [1]{``#1''}%
\providecommand \bibnamefont  [1]{#1}%
\providecommand \bibfnamefont [1]{#1}%
\providecommand \citenamefont [1]{#1}%
\providecommand \href@noop [0]{\@secondoftwo}%
\providecommand \href [0]{\begingroup \@sanitize@url \@href}%
\providecommand \@href[1]{\@@startlink{#1}\@@href}%
\providecommand \@@href[1]{\endgroup#1\@@endlink}%
\providecommand \@sanitize@url [0]{\catcode `\\12\catcode `\$12\catcode
  `\&12\catcode `\#12\catcode `\^12\catcode `\_12\catcode `\%12\relax}%
\providecommand \@@startlink[1]{}%
\providecommand \@@endlink[0]{}%
\providecommand \url  [0]{\begingroup\@sanitize@url \@url }%
\providecommand \@url [1]{\endgroup\@href {#1}{\urlprefix }}%
\providecommand \urlprefix  [0]{URL }%
\providecommand \Eprint [0]{\href }%
\providecommand \doibase [0]{https://doi.org/}%
\providecommand \selectlanguage [0]{\@gobble}%
\providecommand \bibinfo  [0]{\@secondoftwo}%
\providecommand \bibfield  [0]{\@secondoftwo}%
\providecommand \translation [1]{[#1]}%
\providecommand \BibitemOpen [0]{}%
\providecommand \bibitemStop [0]{}%
\providecommand \bibitemNoStop [0]{.\EOS\space}%
\providecommand \EOS [0]{\spacefactor3000\relax}%
\providecommand \BibitemShut  [1]{\csname bibitem#1\endcsname}%
\let\auto@bib@innerbib\@empty
%</preamble>
\bibitem [{\citenamefont {Einstein}\ and\ \citenamefont
  {Rosen}(1937)}]{EinsteinRosen37}%
  \BibitemOpen
  \bibfield  {author} {\bibinfo {author} {\bibfnamefont {A.}~\bibnamefont
  {Einstein}}\ and\ \bibinfo {author} {\bibfnamefont {N.}~\bibnamefont
  {Rosen}},\ }\href {https://doi.org/10.1016/S0016-0032(37)90583-0} {\bibfield
  {journal} {\bibinfo  {journal} {J. Frankl.\ Inst.}\ }\textbf {\bibinfo
  {volume} {223}},\ \bibinfo {pages} {43} (\bibinfo {year} {1937})}\BibitemShut
  {NoStop}%
\bibitem [{\citenamefont {Piran}\ and\ \citenamefont
  {Safier}(1985)}]{PiranSafier:1985}%
  \BibitemOpen
  \bibfield  {author} {\bibinfo {author} {\bibfnamefont {T.}~\bibnamefont
  {Piran}}\ and\ \bibinfo {author} {\bibfnamefont {P.~N.}\ \bibnamefont
  {Safier}},\ }\href {https://doi.org/10.1038/318271a0} {\bibfield  {journal}
  {\bibinfo  {journal} {Nature (London)}\ }\textbf {\bibinfo {volume} {318}},\
  \bibinfo {pages} {271} (\bibinfo {year} {1985})}\BibitemShut {NoStop}%
\bibitem [{\citenamefont {Marder}(1958)}]{marder}%
  \BibitemOpen
  \bibfield  {author} {\bibinfo {author} {\bibfnamefont {L.}~\bibnamefont
  {Marder}},\ }\href {https://doi.org/10.1098/rspa.1958.0058} {\bibfield
  {journal} {\bibinfo  {journal} {Proc.\ R.\ Soc.\ A Math.}\ }\textbf {\bibinfo
  {volume} {244}},\ \bibinfo {pages} {524 } (\bibinfo {year}
  {1958})}\BibitemShut {NoStop}%
\bibitem [{\citenamefont {Bonnor}(1957)}]{bpulse}%
  \BibitemOpen
  \bibfield  {author} {\bibinfo {author} {\bibfnamefont {W.~B.}\ \bibnamefont
  {Bonnor}},\ }\href {https://www.jstor.org/stable/24900449} {\bibfield
  {journal} {\bibinfo  {journal} {J. Math. Mech.}\ }\textbf {\bibinfo {volume}
  {6}},\ \bibinfo {pages} {203} (\bibinfo {year} {1957})}\BibitemShut {NoStop}%
\bibitem [{\citenamefont {Stephani}(2003)}]{Stephani:2003}%
  \BibitemOpen
  \bibfield  {author} {\bibinfo {author} {\bibfnamefont {H.}~\bibnamefont
  {Stephani}},\ }\href {https://doi.org/10.1023/A:1022330218708} {\bibfield
  {journal} {\bibinfo  {journal} {Gen.\ Relativ.\ Gravit.}\ }\textbf {\bibinfo
  {volume} {35}},\ \bibinfo {pages} {467} (\bibinfo {year} {2003})}\BibitemShut
  {NoStop}%
\bibitem [{\citenamefont {Halilsoy}(1988)}]{Halilsoy:1988}%
  \BibitemOpen
  \bibfield  {author} {\bibinfo {author} {\bibfnamefont {M.}~\bibnamefont
  {Halilsoy}},\ }\href {https://doi.org/10.1007/BF02725615} {\bibfield
  {journal} {\bibinfo  {journal} {Nuovo Cimento Soc. Ital. Fis. B}\ }\textbf
  {\bibinfo {volume} {102}},\ \bibinfo {pages} {563} (\bibinfo {year}
  {1988})}\BibitemShut {NoStop}%
\bibitem [{\citenamefont {Chandrasekhar}(1986)}]{chandra}%
  \BibitemOpen
  \bibfield  {author} {\bibinfo {author} {\bibfnamefont {S.}~\bibnamefont
  {Chandrasekhar}},\ }\href {https://doi.org/10.1098/rspa.1986.0117} {\bibfield
   {journal} {\bibinfo  {journal} {Proc.\ R.\ Soc.\ A Math.}\ }\textbf
  {\bibinfo {volume} {408}},\ \bibinfo {pages} {209} (\bibinfo {year}
  {1986})}\BibitemShut {NoStop}%
\bibitem [{\citenamefont {Nikiel}\ and\ \citenamefont
  {Szybka}(2025)}]{NikielSzybka:2025}%
  \BibitemOpen
  \bibfield  {author} {\bibinfo {author} {\bibfnamefont {K.}~\bibnamefont
  {Nikiel}}\ and\ \bibinfo {author} {\bibfnamefont {S.~J.}\ \bibnamefont
  {Szybka}},\ }\href {https://doi.org/10.1103/PhysRevD.111.104015} {\bibfield
  {journal} {\bibinfo  {journal} {Phys. Rev. D}\ }\textbf {\bibinfo {volume}
  {111}},\ \bibinfo {pages} {104015} (\bibinfo {year} {2025})}\BibitemShut
  {NoStop}%
\bibitem [{\citenamefont {Green}\ and\ \citenamefont {Wald}(2011)}]{greenwald}%
  \BibitemOpen
  \bibfield  {author} {\bibinfo {author} {\bibfnamefont {S.~R.}\ \bibnamefont
  {Green}}\ and\ \bibinfo {author} {\bibfnamefont {R.~M.}\ \bibnamefont
  {Wald}},\ }\href {https://doi.org/10.1103/PhysRevD.83.084020} {\bibfield
  {journal} {\bibinfo  {journal} {Phys.\ Rev.\ D}\ }\textbf {\bibinfo {volume}
  {83}},\ \bibinfo {pages} {084020} (\bibinfo {year} {2011})}\BibitemShut
  {NoStop}%
\bibitem [{\citenamefont {Isaacson}(1968{\natexlab{a}})}]{isaacson1}%
  \BibitemOpen
  \bibfield  {author} {\bibinfo {author} {\bibfnamefont {R.~A.}\ \bibnamefont
  {Isaacson}},\ }\href {https://doi.org/10.1103/PhysRev.166.1263} {\bibfield
  {journal} {\bibinfo  {journal} {Phys. Rev.}\ }\textbf {\bibinfo {volume}
  {166}},\ \bibinfo {pages} {1263} (\bibinfo {year}
  {1968}{\natexlab{a}})}\BibitemShut {NoStop}%
\bibitem [{\citenamefont {Isaacson}(1968{\natexlab{b}})}]{isaacson2}%
  \BibitemOpen
  \bibfield  {author} {\bibinfo {author} {\bibfnamefont {R.~A.}\ \bibnamefont
  {Isaacson}},\ }\href {https://doi.org/10.1103/PhysRev.166.1272} {\bibfield
  {journal} {\bibinfo  {journal} {Phys. Rev.}\ }\textbf {\bibinfo {volume}
  {166}},\ \bibinfo {pages} {1272} (\bibinfo {year}
  {1968}{\natexlab{b}})}\BibitemShut {NoStop}%
\bibitem [{\citenamefont {Maccallum}\ and\ \citenamefont
  {Taub}(1973)}]{Maccallum:1973gf}%
  \BibitemOpen
  \bibfield  {author} {\bibinfo {author} {\bibfnamefont {M.~A.~H.}\
  \bibnamefont {Maccallum}}\ and\ \bibinfo {author} {\bibfnamefont {A.~H.}\
  \bibnamefont {Taub}},\ }\href {https://doi.org/10.1007/BF01645977} {\bibfield
   {journal} {\bibinfo  {journal} {Commun. Math. Phys.}\ }\textbf {\bibinfo
  {volume} {30}},\ \bibinfo {pages} {153} (\bibinfo {year} {1973})}\BibitemShut
  {NoStop}%
\bibitem [{\citenamefont {{Burnett}}(1989)}]{burnett}%
  \BibitemOpen
  \bibfield  {author} {\bibinfo {author} {\bibfnamefont {G.~A.}\ \bibnamefont
  {{Burnett}}},\ }\href {https://doi.org/10.1063/1.528594} {\bibfield
  {journal} {\bibinfo  {journal} {J. Math. Phys.}\ }\textbf {\bibinfo {volume}
  {30}},\ \bibinfo {pages} {90} (\bibinfo {year} {1989})}\BibitemShut {NoStop}%
\bibitem [{\citenamefont {Podolský}\ and\ \citenamefont
  {Svítek}(2004)}]{Podolsky:2003bm}%
  \BibitemOpen
  \bibfield  {author} {\bibinfo {author} {\bibfnamefont {J.}~\bibnamefont
  {Podolský}}\ and\ \bibinfo {author} {\bibfnamefont {O.}~\bibnamefont
  {Svítek}},\ }\href {https://doi.org/10.1023/B:GERG.0000010483.02257.90}
  {\bibfield  {journal} {\bibinfo  {journal} {Gen. Rel. Grav.}\ }\textbf
  {\bibinfo {volume} {36}},\ \bibinfo {pages} {387} (\bibinfo {year}
  {2004})}\BibitemShut {NoStop}%
\bibitem [{\citenamefont {Huneau}\ and\ \citenamefont
  {Luk}(2025)}]{Huneau:2025mry}%
  \BibitemOpen
  \bibfield  {author} {\bibinfo {author} {\bibfnamefont {C.}~\bibnamefont
  {Huneau}}\ and\ \bibinfo {author} {\bibfnamefont {J.}~\bibnamefont {Luk}},\
  }\href@noop {} {} (\bibinfo {year} {2025}),\ \Eprint
  {https://arxiv.org/abs/2506.21779} {arXiv:2506.21779 [gr-qc]} \BibitemShut
  {NoStop}%
\bibitem [{\citenamefont {Huneau}\ and\ \citenamefont
  {Luk}(2024{\natexlab{a}})}]{Huneau:2024}%
  \BibitemOpen
  \bibfield  {author} {\bibinfo {author} {\bibfnamefont {C.}~\bibnamefont
  {Huneau}}\ and\ \bibinfo {author} {\bibfnamefont {J.}~\bibnamefont {Luk}},\
  }\href {https://doi.org/10.1088/1361-6382/ad5487} {\bibfield  {journal}
  {\bibinfo  {journal} {Class. Quant. Grav.}\ }\textbf {\bibinfo {volume}
  {41}},\ \bibinfo {pages} {143002} (\bibinfo {year}
  {2024}{\natexlab{a}})}\BibitemShut {NoStop}%
\bibitem [{\citenamefont {Morgan}(1973)}]{morgan}%
  \BibitemOpen
  \bibfield  {author} {\bibinfo {author} {\bibfnamefont {T.~A.}\ \bibnamefont
  {Morgan}},\ }\href {https://doi.org/10.1007/BF00759846} {\bibfield  {journal}
  {\bibinfo  {journal} {Gen. Relativ. Gravit.}\ }\textbf {\bibinfo {volume}
  {4}},\ \bibinfo {pages} {273} (\bibinfo {year} {1973})}\BibitemShut {NoStop}%
\bibitem [{\citenamefont {Choquet-Bruhat}(1969)}]{choquet}%
  \BibitemOpen
  \bibfield  {author} {\bibinfo {author} {\bibfnamefont {Y.}~\bibnamefont
  {Choquet-Bruhat}},\ }\href {https://doi.org/10.1007/BF01646432} {\bibfield
  {journal} {\bibinfo  {journal} {Commun. Math. Phys.}\ }\textbf {\bibinfo
  {volume} {12}},\ \bibinfo {pages} {16} (\bibinfo {year} {1969})}\BibitemShut
  {NoStop}%
\bibitem [{\citenamefont {Huneau}\ and\ \citenamefont
  {Luk}(2024{\natexlab{b}})}]{Huneau:2024lrb}%
  \BibitemOpen
  \bibfield  {author} {\bibinfo {author} {\bibfnamefont {C.}~\bibnamefont
  {Huneau}}\ and\ \bibinfo {author} {\bibfnamefont {J.}~\bibnamefont {Luk}},\
  }\href@noop {} {} (\bibinfo {year} {2024}{\natexlab{b}}),\ \Eprint
  {https://arxiv.org/abs/2403.03470} {arXiv:2403.03470 [gr-qc]} \BibitemShut
  {NoStop}%
\bibitem [{\citenamefont {Green}\ and\ \citenamefont
  {Wald}(2013)}]{greenwald2}%
  \BibitemOpen
  \bibfield  {author} {\bibinfo {author} {\bibfnamefont {S.~R.}\ \bibnamefont
  {Green}}\ and\ \bibinfo {author} {\bibfnamefont {R.~M.}\ \bibnamefont
  {Wald}},\ }\href {https://doi.org/10.1103/PhysRevD.87.124037} {\bibfield
  {journal} {\bibinfo  {journal} {Phys. Rev. D}\ }\textbf {\bibinfo {volume}
  {87}},\ \bibinfo {pages} {124037} (\bibinfo {year} {2013})}\BibitemShut
  {NoStop}%
\bibitem [{\citenamefont {Szybka}\ \emph {et~al.}(2014)\citenamefont {Szybka},
  \citenamefont {G\l{}\'od}, \citenamefont {Wyr\ifmmode~\mbox{\k{e}}\else
  \k{e}\fi{}bowski},\ and\ \citenamefont {Konieczny}}]{ourwm}%
  \BibitemOpen
  \bibfield  {author} {\bibinfo {author} {\bibfnamefont {S.~J.}\ \bibnamefont
  {Szybka}}, \bibinfo {author} {\bibfnamefont {K.}~\bibnamefont {G\l{}\'od}},
  \bibinfo {author} {\bibfnamefont {M.~J.}\ \bibnamefont
  {Wyr\ifmmode~\mbox{\k{e}}\else \k{e}\fi{}bowski}},\ and\ \bibinfo {author}
  {\bibfnamefont {A.}~\bibnamefont {Konieczny}},\ }\href
  {https://doi.org/10.1103/PhysRevD.89.044033} {\bibfield  {journal} {\bibinfo
  {journal} {Phys. Rev. D}\ }\textbf {\bibinfo {volume} {89}},\ \bibinfo
  {pages} {044033} (\bibinfo {year} {2014})}\BibitemShut {NoStop}%
\bibitem [{\citenamefont {Szybka}\ and\ \citenamefont
  {Wyr\ifmmode~\mbox{\k{e}}\else \k{e}\fi{}bowski}(2016)}]{ers}%
  \BibitemOpen
  \bibfield  {author} {\bibinfo {author} {\bibfnamefont {S.~J.}\ \bibnamefont
  {Szybka}}\ and\ \bibinfo {author} {\bibfnamefont {M.~J.}\ \bibnamefont
  {Wyr\ifmmode~\mbox{\k{e}}\else \k{e}\fi{}bowski}},\ }\href
  {https://doi.org/10.1103/PhysRevD.94.024059} {\bibfield  {journal} {\bibinfo
  {journal} {Phys.\ Rev.\ D}\ }\textbf {\bibinfo {volume} {94}},\ \bibinfo
  {pages} {024059} (\bibinfo {year} {2016})}\BibitemShut {NoStop}%
\bibitem [{\citenamefont {Huneau}\ and\ \citenamefont
  {Luk}(2018)}]{Huneau:2017led}%
  \BibitemOpen
  \bibfield  {author} {\bibinfo {author} {\bibfnamefont {C.}~\bibnamefont
  {Huneau}}\ and\ \bibinfo {author} {\bibfnamefont {J.}~\bibnamefont {Luk}},\
  }\href {https://doi.org/10.1215/00127094-2018-0035} {\bibfield  {journal}
  {\bibinfo  {journal} {Duke Math. J.}\ }\textbf {\bibinfo {volume} {167}},\
  \bibinfo {pages} {3315} (\bibinfo {year} {2018})}\BibitemShut {NoStop}%
%%CITATION = ARXIV:1706.09501;%%
\bibitem [{\citenamefont {Luk}\ and\ \citenamefont
  {Rodnianski}(2020)}]{Luk:2020pyn}%
  \BibitemOpen
  \bibfield  {author} {\bibinfo {author} {\bibfnamefont {J.}~\bibnamefont
  {Luk}}\ and\ \bibinfo {author} {\bibfnamefont {I.}~\bibnamefont
  {Rodnianski}},\ }\href@noop {} {} (\bibinfo {year} {2020}),\ \Eprint
  {https://arxiv.org/abs/2009.08968} {arXiv:2009.08968 [math.AP]} \BibitemShut
  {NoStop}%
\bibitem [{\citenamefont {Caves}\ \emph {et~al.}(1980)\citenamefont {Caves},
  \citenamefont {Thorne}, \citenamefont {Drever}, \citenamefont {Sandberg},\
  and\ \citenamefont {Zimmermann}}]{RevModPhys.52.341}%
  \BibitemOpen
  \bibfield  {author} {\bibinfo {author} {\bibfnamefont {C.~M.}\ \bibnamefont
  {Caves}}, \bibinfo {author} {\bibfnamefont {K.~S.}\ \bibnamefont {Thorne}},
  \bibinfo {author} {\bibfnamefont {R.~W.~P.}\ \bibnamefont {Drever}}, \bibinfo
  {author} {\bibfnamefont {V.~D.}\ \bibnamefont {Sandberg}},\ and\ \bibinfo
  {author} {\bibfnamefont {M.}~\bibnamefont {Zimmermann}},\ }\href
  {https://doi.org/10.1103/RevModPhys.52.341} {\bibfield  {journal} {\bibinfo
  {journal} {Rev. Mod. Phys.}\ }\textbf {\bibinfo {volume} {52}},\ \bibinfo
  {pages} {341} (\bibinfo {year} {1980})}\BibitemShut {NoStop}%
\bibitem [{\citenamefont {{Grishchuk}}\ and\ \citenamefont
  {{Sazhin}}(1975)}]{GrishchukSazhin1975}%
  \BibitemOpen
  \bibfield  {author} {\bibinfo {author} {\bibfnamefont {L.~P.}\ \bibnamefont
  {{Grishchuk}}}\ and\ \bibinfo {author} {\bibfnamefont {M.~V.}\ \bibnamefont
  {{Sazhin}}},\ }\href {https://jetp.ras.ru/cgi-bin/dn/e_041_05_0787.pdf}
  {\bibfield  {journal} {\bibinfo  {journal} {Z. Eksp.\ Teor.\ Fiz.}\ }\textbf
  {\bibinfo {volume} {68}},\ \bibinfo {pages} {1569} (\bibinfo {year}
  {1975})}\BibitemShut {NoStop}%
\bibitem [{\citenamefont {Thorne}(1965)}]{thorne}%
  \BibitemOpen
  \bibfield  {author} {\bibinfo {author} {\bibfnamefont {K.~S.}\ \bibnamefont
  {Thorne}},\ }\href {https://doi.org/10.1103/PhysRev.138.B251} {\bibfield
  {journal} {\bibinfo  {journal} {Phys.\ Rev.}\ }\textbf {\bibinfo {volume}
  {138}},\ \bibinfo {pages} {B251} (\bibinfo {year} {1965})}\BibitemShut
  {NoStop}%
\bibitem [{\citenamefont {Ernst}(1968{\natexlab{a}})}]{Ernst:1967wx}%
  \BibitemOpen
  \bibfield  {author} {\bibinfo {author} {\bibfnamefont {F.~J.}\ \bibnamefont
  {Ernst}},\ }\href {https://doi.org/10.1103/PhysRev.167.1175} {\bibfield
  {journal} {\bibinfo  {journal} {Phys. Rev.}\ }\textbf {\bibinfo {volume}
  {167}},\ \bibinfo {pages} {1175} (\bibinfo {year}
  {1968}{\natexlab{a}})}\BibitemShut {NoStop}%
\bibitem [{\citenamefont {Ernst}(1968{\natexlab{b}})}]{Ernst:1967by}%
  \BibitemOpen
  \bibfield  {author} {\bibinfo {author} {\bibfnamefont {F.~J.}\ \bibnamefont
  {Ernst}},\ }\href {https://doi.org/10.1103/PhysRev.168.1415} {\bibfield
  {journal} {\bibinfo  {journal} {Phys. Rev.}\ }\textbf {\bibinfo {volume}
  {168}},\ \bibinfo {pages} {1415} (\bibinfo {year}
  {1968}{\natexlab{b}})}\BibitemShut {NoStop}%
\bibitem [{\citenamefont {Szybka}\ and\ \citenamefont
  {Naqvi}(2023)}]{szybkanaqvi}%
  \BibitemOpen
  \bibfield  {author} {\bibinfo {author} {\bibfnamefont {S.~J.}\ \bibnamefont
  {Szybka}}\ and\ \bibinfo {author} {\bibfnamefont {S.~U.}\ \bibnamefont
  {Naqvi}},\ }\href {https://doi.org/10.1103/PhysRevD.108.L081501} {\bibfield
  {journal} {\bibinfo  {journal} {Phys. Rev. D}\ }\textbf {\bibinfo {volume}
  {108}} (\bibinfo {year} {2023})}\BibitemShut {NoStop}%
\bibitem [{\citenamefont {Szybka}\ and\ \citenamefont
  {Cieślik}(2019)}]{Szybka:2019}%
  \BibitemOpen
  \bibfield  {author} {\bibinfo {author} {\bibfnamefont {S.~J.}\ \bibnamefont
  {Szybka}}\ and\ \bibinfo {author} {\bibfnamefont {A.}~\bibnamefont
  {Cieślik}},\ }\href {https://doi.org/10.1103/PhysRevD.100.064025} {\bibfield
   {journal} {\bibinfo  {journal} {Phys.\ Rev.\ D}\ }\textbf {\bibinfo {volume}
  {100}},\ \bibinfo {pages} {064025} (\bibinfo {year} {2019})}\BibitemShut
  {NoStop}%
%%CITATION = ARXIV:1901.10285;%%
\bibitem [{\citenamefont {Stachel}(1966)}]{Stachel1966}%
  \BibitemOpen
  \bibfield  {author} {\bibinfo {author} {\bibfnamefont {J.~J.}\ \bibnamefont
  {Stachel}},\ }\href {https://doi.org/10.1063/1.1705036} {\bibfield  {journal}
  {\bibinfo  {journal} {J. Math.\ Phys.}\ }\textbf {\bibinfo {volume} {7}},\
  \bibinfo {pages} {1321} (\bibinfo {year} {1966})}\BibitemShut {NoStop}%
\bibitem [{\citenamefont {Chru\ifmmode~\acute{s}\else \'{s}\fi{}ciel}\ \emph
  {et~al.}(2012)\citenamefont {Chru\ifmmode~\acute{s}\else \'{s}\fi{}ciel},
  \citenamefont {\"Olz},\ and\ \citenamefont {Szybka}}]{diagrammatics}%
  \BibitemOpen
  \bibfield  {author} {\bibinfo {author} {\bibfnamefont {P.~T.}\ \bibnamefont
  {Chru\ifmmode~\acute{s}\else \'{s}\fi{}ciel}}, \bibinfo {author}
  {\bibfnamefont {C.~R.}\ \bibnamefont {\"Olz}},\ and\ \bibinfo {author}
  {\bibfnamefont {S.~J.}\ \bibnamefont {Szybka}},\ }\href
  {https://doi.org/10.1103/PhysRevD.86.124041} {\bibfield  {journal} {\bibinfo
  {journal} {Phys. Rev. D}\ }\textbf {\bibinfo {volume} {86}},\ \bibinfo
  {pages} {124041} (\bibinfo {year} {2012})}\BibitemShut {NoStop}%
\bibitem [{\citenamefont {Mart\'{\i}n-Garc\'{\i}a}(2021)}]{xAct}%
  \BibitemOpen
  \bibfield  {author} {\bibinfo {author} {\bibfnamefont {J.~M.}\ \bibnamefont
  {Mart\'{\i}n-Garc\'{\i}a}},\ }\href {http://www.xact.es} {\bibinfo {title}
  {\href{http://www.xact.es}{{xAct}: Efficient Tensor Computer Algebra}}}
  (\bibinfo {year} {2021})\BibitemShut {NoStop}%
\bibitem [{\citenamefont {Ölz}\ and\ \citenamefont {Szybka}(2013)}]{Olz}%
  \BibitemOpen
  \bibfield  {author} {\bibinfo {author} {\bibfnamefont {C.~R.}\ \bibnamefont
  {Ölz}}\ and\ \bibinfo {author} {\bibfnamefont {S.~J.}\ \bibnamefont
  {Szybka}},\ }\href {https://arxiv.org/abs/1305.2177} {} (\bibinfo {year}
  {2013}),\ \Eprint {https://arxiv.org/abs/1305.2177} {arXiv:1305.2177 [gr-qc]}
  \BibitemShut {NoStop}%
\end{thebibliography}%

\end{document}